\documentclass[%
 twocolumn, 
 amsmath,amssymb,
 aps,
prb,
floatfix,
]{revtex4-1}
\usepackage{graphicx}
\usepackage{dcolumn}
\usepackage{bm}
\usepackage{color} 
\usepackage{wrapfig}
\usepackage{lipsum} 

\usepackage[normalem]{ulem}
\usepackage[usenames, dvipsnames]{xcolor}
\newcommand\blue[1]{{\color{blue}#1}}

\definecolor{green}{rgb}{0.0, 0.5, 0.0}

\begin{document}
\preprint{APS/123-QED}
 
\title{Magnetic properties of bilayer VI$_{3}$:\\
Role of trigonal crystal field and electric-field tuning } 



\author{Thi Phuong Thao Nguyen}
\author{Kunihiko Yamauchi}
\altaffiliation{current address: Elements Strategy Initiative for Catalysts and Batteries (ESICB), Kyoto University, Kyoto 615-8245, Japan}
\author{Tamio Oguchi}
\affiliation{%
 Institute of Scientific and Industrial Research, Osaka University, 8-1 Mihogaoka, Ibaraki, Osaka, 567-0047, Japan
}%
\author{Danila Amoroso}
\author{Silvia Picozzi}
\affiliation{%
Consiglio Nazionale delle Ricerche (CNR-SPIN), Unit\`a di Ricerca presso Terzi c/o Universit\`a ``G. D'Annunzio", 66100 Chiety, Italy
}%

\date{\today}

\begin{abstract}

The  magnetic properties 
of  two-dimensional VI$_3$ bilayer are the focus of our  first-principles  analysis, highlighting the role of  trigonal
crystal-field effects and carried out in comparison with the CrI$_3$ prototypical case, where the effects are absent. 
In VI$_3$ bilayers, 
 the empty $a_{1g}$ state - consistent with the observed trigonal distortion - is found to play a crucial role in both stabilizing the insulating state and in determining the inter-layer magnetic interaction. Indeed, an analysis based
on maximally-localized Wannier functions 
allows to 
evaluate the interlayer exchange interactions in two different VI$_3$ stackings (labelled AB and AB'), to interpret the results in terms of virtual-hopping mechanism,  and to highlight the strongest  hopping  channels underlying the magnetic interlayer coupling. 
Upon application of  electric fields perpendicular to the slab, we find that the magnetic ground-state in the AB' stacking can be switched from antiferromagnetic to ferromagnetic, suggesting VI$_3$ bilayer as an appealing candidate for electric-field-driven miniaturized spintronic devices.
\end{abstract}


\maketitle


\section{Introduction}
Boosted by the experimental discovery of intrinsic magnetism in atomically thin layers of CrI$_3$, Cr$_2$Ge$_2$Te$_6$ and Fe$_3$GeTe$_2$, \cite{cri3,crgete, fe3gete2}
two-dimensional (2D) van der Waals (vdW) magnets have recently received an increasing attention.
 Interestingly, the control of 2D magnetism in few atomic layers is enabled by external electric field \cite{cri3_field1, cri3_field2, cri3_field3} or by electron-hole doping,\cite{cri3_doping,fe3gete2} making them particularly appealing for potential spintronic applications.

Among those materials, VI$_3$,  belonging to the family of transition-metal (M) trihalides MX$_3$ (X = Cl, B, and I) with honeycomb arrangement of the metal cations (similar to the most studied CrI$_3 $ \cite{cri3, cri3_field2, cri3_doping, cri3_struc, cri3_stack1, cri3_stack2, cri3_stack3, cri3_mae, cri3_ligand, cri3_field1, cri3_field3}), has recently emerged as a potential 2D ferromagnet~\cite{vi3_bulk2_insu, vi3_bulk3_insu, vi3_ml1_2states, vi3_ml2_metallic, vi3_ml3_insu_soc, vi3_bulk_ml_metallic, vi3_bulk4_metallic}.  
It is known since more than 30 years ago that, in the bulk form, 
VI$_3$ becomes ferromagnetic below a Curie temperature of $\it{T}_c\simeq$ 55~K, similar to CrI$_3$ (with $\it{T}_c\simeq$ 68~K)~\cite{vi3_bulk1}. Conversely, the structural properties are still under debate. Experimental characterizations of the crystal structure have in fact reported that VI$_3$ undergoes a structural phase transition around $79$~K, changing its symmetry across still unclear phases:
the high-temperature (HT) crystal structure was proposed to be either trigonal {\it P}31{\it c}~\cite{vi3_bulk2_insu} or rhombohedral {\it R}3 \cite{vi3_bulk4_metallic}, or monoclinic {\it C}2/{\it m} structure \cite{vi3_bulk3_insu}; the low-temperature (LT) crystal structure was proposed to be either {\it C}2/{\it c} \cite{vi3_bulk2_insu} or {\it R}\=3 structure \cite{vi3_bulk3_insu}. 
Optical and electrical transport measurements have clearly showed bulk VI$_3$ to be a semiconductor with an optical band gap of $\sim 0.67$ eV\cite{vi3_bulk2_insu}. 
However, from the theoretical point of view, the understanding and modeling of the electronic properties are the focus of present debate. Some studies have in fact reported that bulk VI$_3$ is a Mott-insulator with a bandgap of about 1 eV\cite{vi3_bulk2_insu, vi3_bulk3_insu}, whereas  
others have claimed a half-metallic character\cite{vi3_bulk_ml_metallic, vi3_bulk4_metallic}.

In a thin-film limit, to the best of our knowledge, no experimental studies have been  reported for  atomic layers of VI$_3$. On the theoretical side, current  studies are  controversial with respect to the electronic properties, in analogy to the situation for the corresponding bulk phase. 
In particular, by analysing electronic properties in the VI$_3$ monolayer, in Ref.\cite{vi3_ml1_2states}  a Mott-insulator ground state is proposed, reported to be lower in energy than the half-metallic state (by $\sim$0.3 eV/f.u). 
Other attempts to explain the Mott-insulator ground state have also been reported:  the authors in Ref.\cite{vi3_ml2_metallic} have proposed orbital-ordered phases accompanying the lattice distortion,  
while the authors of Ref.\cite{vi3_ml3_insu_soc} ascribed the gap opening to combined effects of spin-orbit coupling (SOC) and Hubbard $U$ correlations. 
On the other hand, a consensus is reached with respect to the magnetic properties of a VI$_3$ single layer: current theoretical characterizations report a ferromagnetic (FM) exchange coupling between first V-site neighbors, {\em i.e.} FM intra-layer coupling. The inter-layer magnetic stability has  also been investigated for bilayer VI$_3$. In particular in ~Ref.\cite{vi3_ml1_2states} it is reported that the inter-layer magnetic stability is sensitive to the layer stacking, in line with 
previous works on bilayer CrI$_3$ \cite{cri3_stack1, cri3_stack2, cri3_stack3}. For the sake of completeness, we mention that in Ref.~\cite{vi3_stack} it is claimed  bilayer VI$_3$ to show a stacking-independent ferromagnetic ground state, but considering the half-metallic state rather than the Mott-insulating one. 

A deep understanding of the VI$_3$ electronic structure is needed to interpret the related magnetic properties, both in the monolayer and in the bilayer case.  
In this study, we  therefore
focus on the crucial role of trigonal crystal field effects in determining the VI$_3$ insulating behaviour and the related magnetic properties. As a counterexample, we consider the prototypical 2D magnet, {\em i.e.} CrI$_3$ (where trigonal crystal field effects are absent), and we carry out a one-to-one comparison between VI$_3$ and CrI$_3$. In particular, we first focus on the monolayer and discuss crystal field effects (Section III A) and density of states (Section III B). Then we move to the 
the magnetic properties of bilayer halides, by considering two different stacking arrangements. We concentrate on the interlayer exchange coupling, interpreting it in terms of virtual hopping mechanisms, highlighting the most efficient hopping paths (Section IV A) and addressing the effects of an external  electric field in tuning the magnetic stability (Section IV B). Finally, we draw our  conclusions in Section V.





\section{Computational method}
Density-functional theory (DFT) calculations were performed using the VASP code \cite{vasp} within the generalized gradient approximation (GGA) \cite{gga}. The van der Waals (vdW) interactions were included for bilayer structure calculations. The rotationally invariant GGA + \textit{U} method was employed to account for correlation effects \cite{gga+u}. On-site Coulomb interaction for transition-metal 3\textit{d} orbitals was considered with an effective \textit{U} 
of 2.0 eV \cite{cri3_field2, Liechtenstein.LDAU}. 
 Semiconducting and metallic states are initialized 
 via the density matrix within the GGA+$U$ scheme \cite{Liechtenstein.LDAU}. 
Brillouin zone integrations were performed using a \textit{k}-point grid of 6 $\times$ 6$\times$ 1 for the structure optimization. 
Band structures and density of states (also including electric fields) were 
calculated by using 12 $\times$ 12 $\times$ 1 \textit{k}-points mesh.
Electric fields are applied perpendicularly to the surface by saw-tooth-like potential with dipole correction\cite{dipole}. 
The maximally-localized Wannier functions (MLWFs) were calculated by using WANNIER90 tool~\cite{wannier90} interfaced with the VASP code.


\section{Results for Monolayer VI$_3$}

\subsection{Crystal-field effects}

\begin{figure}[!htbp]
\begin{center}
{
\includegraphics[width=80mm, angle=0]{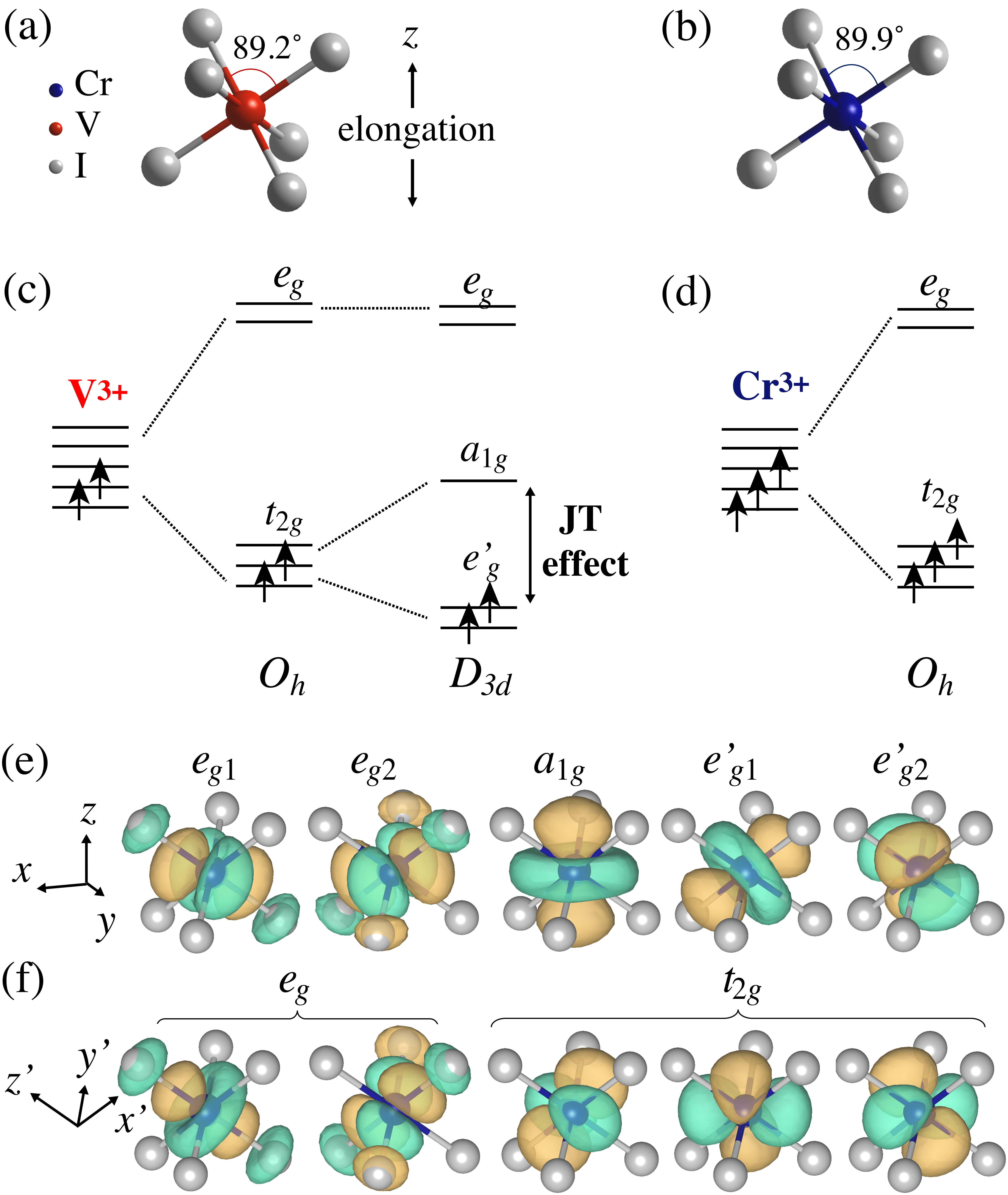}
}
\caption {\label{fig:JT}
\color{black}
Distortion of 
the crystal structure in (a) VI$_3$ and (b) CrI$_3$. The distortion does not change the bond length but alters the bond angle, leading to a trigonal elongation along the {\it z}-direction. (c) and (d) show the crystal field splitting of $d$ levels in V $3d^2$ and Cr $3d^3$, respectively.
Five  3$d$ Wannier functions reflect the trigonal CEF states in monolayer VI$_3$ (e) and cubic CEF states in monolayer CrI$_3$ (f). 
The isosurface levels of
the Wannier functions were set at $1.5$ $a_0^{-3/2}$ (yellow) and $-1.5$ $a_0^{-3/2}$ (blue), where $a_0$ is the Bohr radius. 
}
\end{center}
\end{figure} 

First we focus on the difference between
VI$_3$ and CrI$_3$ monolayers, as far as crystal field effects are concerned.  
In particular, both VI$_3$ and CrI$_3$ show the magnetic atom coordinated to six I atoms, forming edge-sharing octahedra and resulting in octahedral crystalline electric field (CEF) splitting of the $d$ orbitals into the two-fold $e_g$ and three-fold $t_{2g}$ states. 
The $t_{2g}$ state can be further split into  a doublet \(e'_{g}\) and a singlet $a_{1g}$ state to reduce the band energy, {\em i.e.} a Jahn-Teller (JT) effect in trigonal symmetry
\cite{vi3_ml3_insu_soc, Wu2005}.
This effect leads to the trigonal lattice distortion, but such JT-distortion is not remarkable in this system. 
The I-Cr-I bond angle in CrI$_3$ is almost 90$^{\circ}$ ({\em i.e.} a cubic octahedron), while the I-V-I angle in VI$_3$ approaches 89$^{\circ}$. 
As such, it exhibits a trigonal distortion - elongation along the $z$ axis - still preserving spatial inversion symmetry [see Fig.~\ref{fig:JT}(a) and (b)]. 
Fig.~\ref{fig:JT}(c) and (d) show the difference in CEF splitting for VI$_3$ and CrI$_3$ respectively: as schematically represented, the JT-induced splitting allows for the band gap opening in VI$_3$ by half-filling the majority $e'_{g}$ channel and leaving the $a_{1g}$ empty in the case of V $d^2$. On the other hand, CrI$_3$ is unaffected both because of the Cr $d^3$ valence and the almost cubic CEF.


Within the global Cartesian $\{xyz\}$ coordinate system, Fig.~\ref{fig:JT}(e), the $a_{1g}$ and  $e'_{g}$  states are written in the form \cite{khomskii_book} 
\begin{eqnarray}
\label{eq:t2g}
\left|a_{{\rm 1}g}\right> &=& 3z^2-r^2 \nonumber \\
\left|e'_{{g\rm 1}}\right> &=& \frac{1}{\sqrt{3}} \left (\sqrt{2} \left ( x^2-y^2\right)-zx \right) \nonumber \\
\left|e'_{{g\rm 2}}\right> &=& \frac{1}{\sqrt{3}} \left (\sqrt{2}xy+yz \right)
\end{eqnarray}
where the $z$ axis is parallel to the out-of-layer direction ({\it i.e.} perpendicular to the slab). 
According to the different local crystal field effects, we projected
the Bloch functions onto the local $M$I$_6$ octahedral coordinate system $\{x'y'z'\}$ for CrI$_3$ (with basis axes directed along the Cr-I bonds), and onto the Cartesian $\{xyz\}$ system for VI$_3$ [see Fig.~\ref{fig:JT}(e) and (f)] 

After the projection and maximally localization process, the Wannier functions converged into localized orbitals, as shown in Fig. \ref{fig:JT} (e) and (f): 
orbitals shapes are in agreement with the $e_{g}$-($a_{1g},e'_{g}$) states splitting induced by the trigonal CEF in VI$_3$, and  the $e_{g}$-$t_{2g}$ states induced by the cubic CEF in CrI$_3$. In particular, for the latter, it is possible to recognize the $\left| 3z^2 - r^2\right>$ orbital shape for the $a_{1g}$ state (occupying the empty space at the center of the I$_3$ triangle and pointing along the $z$-direction) and mixed shapes from the ($\left|x^2-y^2\right>,\left|zx\right>$) and ($\left|xy\right>,\left|yz\right>$) orbitals for the $e'_{g1}$ and $e'_{g2}$ states respectively, according to Eq.(\ref{eq:t2g}).    


\subsection{Density of states}


In line with previous works, our DFT calculations on monolayer VI$_3$ converged to two different electronic states, corresponding to half-metallic~\cite{vi3_ml2_metallic} 
and Mott-insulator states \cite{vi3_ml1_2states, vi3_ml3_insu_soc} : 
the two V $d^2$ electrons occupy $t_{2g}$ states as $a^1_{1g}e'^1_{g}$ and $e'^2_{g}a^0_{1g}$, respectively.  
We found that the insulating state (with a direct band gap of $\sim$0.39 eV at the $\Gamma$ point for  $U=2.0$ eV) is lower in energy than the half-metallic state by 1.8 eV/f.u. 
Therefore, we will focus hereafter on the Mott insulator state as the ground state. 
Although Yang {\it et al.} proposed that the insulating ground state of monolayer VI$_3$ is stabilized by  spin-orbit coupling (SOC) splitting rather than by CEF splitting. 
here we remark that the crystal field splitting is energetically dominant over the SOC effect, as usually seen in 3$d$ transition-metal compounds \cite{Goodenough1968, Bruno1989}. 
In fact, when including SOC in our band structure calculations, we observed that it affects the width and energy bands related to the I-$p$ states, but it does not significantly change the V 3$d$ band structure. 
Moreover, as we focus here on the $e'^2_{g}a^0_{1g}$ state, eventual SOC induced splitting would not affect the mechanisms and conclusions presented in this study. 



\begin{figure}[t!]
\begin{center}
\centering
{
\includegraphics[width=60mm, angle=0]{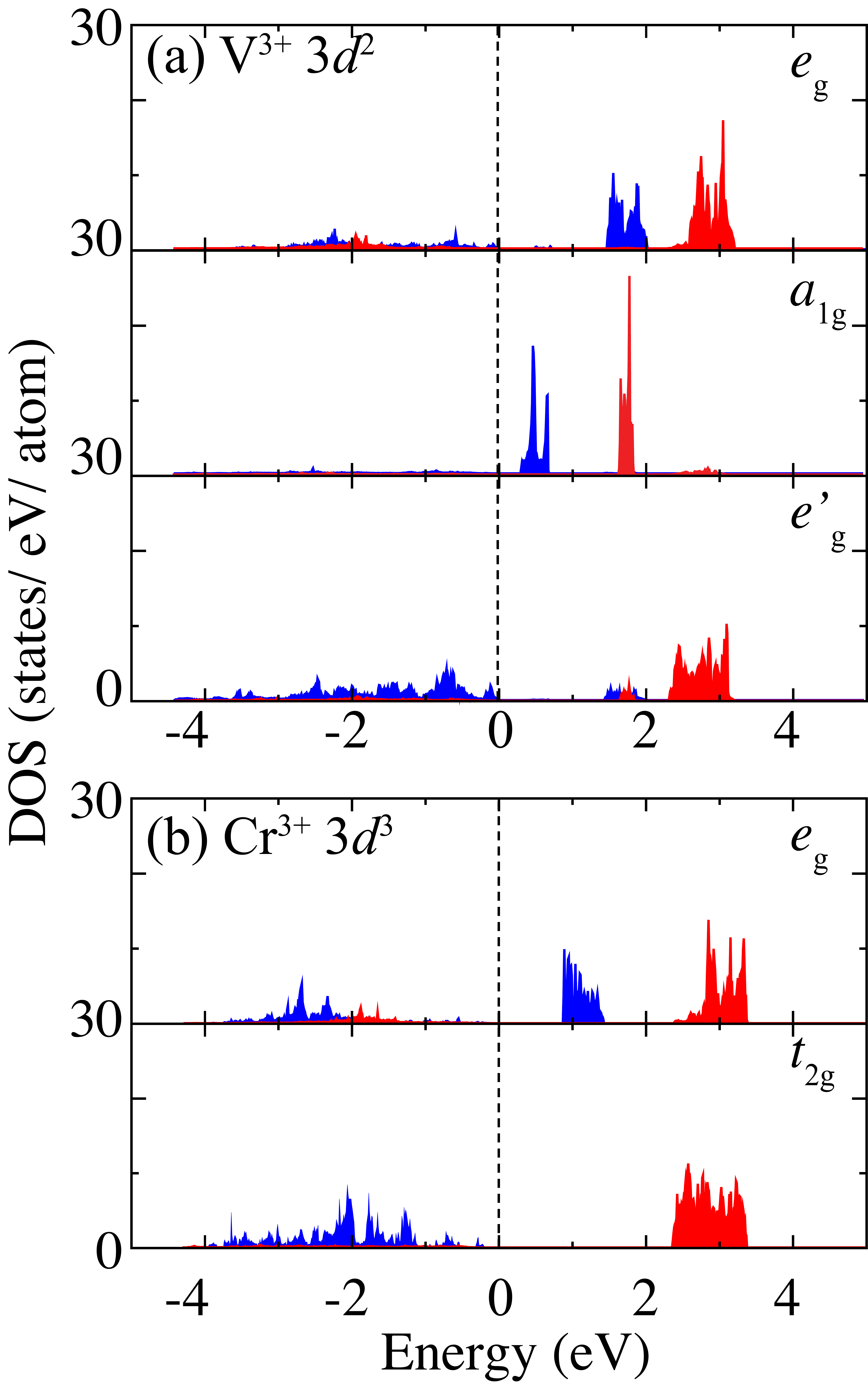}
}
\caption{\label{fig:pdos}
The partial DOS (within $U$ = 2.0 eV) 
projected onto (a) V-3$d$-$D_{3d}$ and (b) Cr-3$d$-$O_h$ CEF states, respectively, via Wannier functions. 
Blue (red) color represents for majority spin (minority spin). Fermi level is set as energy origin. 
}
\end{center}
\end{figure}
 
The partial density of states (DOS) for monolayer CrI$_3$ and VI$_3$ are shown in Fig.~\ref{fig:pdos}; DOS are resolved for each MLWF state, clearly showing the $D_{3d}$ and $O_h$ CEF splitting for VI$_3$ and CrI$_3$ respectively, therefore  validating  our basis functions choice for the Wannier projection as explained in the following. 
 
We note that the Cr $d^3$ partial DOS, projected on the trigonal $D_{3d}$ basic set,  
so as to allow a direct comparison with the V $d^2$ case, shows an overlap in the energy range of the $a_{1g}$ and $e'_{g}$; this reflects the absence of splitting induced by the trigonal CEF, in line with the $O_h$ local octahedral symmetry of CrI$_3$.
In particular, Cr-$d$ orbital states are split into empty $e_g$ and occupied $t_{2g}$ states  with a gap of about 0.9 eV in the majority spin channel (up-spin states); the minority spin channel (down-spin states), unoccupied for both orbital types, does not display any relevant splitting. 
Such a different behavior between the majority and minority spin channels can be ascribed to the $pd$ hybridization: the up-spin $d$-states strongly hybridize  with I-$p$ states located 
at the top of the valence band, 
causing the large CEF splitting supported by the $e_g$-$p$ bonding-antibonding splitting; the down-spin $d$-states are higher in energy, {\it i.e.} away from I-$p$ levels, thus not showing  any significant CEF splitting.  

In VI$_3$, the spin-up channel of the V $d$ states are clearly split into $e_g$, $a_{1g}$, $e_g'$ trigonal CEF states. In particular, the $e'_g$ is the lowest energy state 
with a broad distribution due to the $pd$ hybridization below the Fermi level, similar to  CrI$_3$.
On the other hand, the CEF splitting shows a different behaviour of the empty $d$ states:  
the $a_{1g}$ state becomes the lowest energy state, while the $e'_g$ state still lies in the same energy region as the minority Cr-$t_{2g}$ state. 
This is related to the fact that the
$e'_g$ state has more bonding character with surrounding I $p$ state than the $a_{1g}$ state (compare the orbital shapes in Fig.~\ref{fig:JT}(e)); therefore, the $pd$ hybridization shifts  unoccupied $e'_g$ level up and   occupied I-$p$ level down. 

\section{Results for bilayer VI$_3$}

\subsection{Electronic properties and inter-layer magnetic stability}

\begin{figure}[!htbp]
\begin{center}
{
 \includegraphics[width=70mm, angle=0]{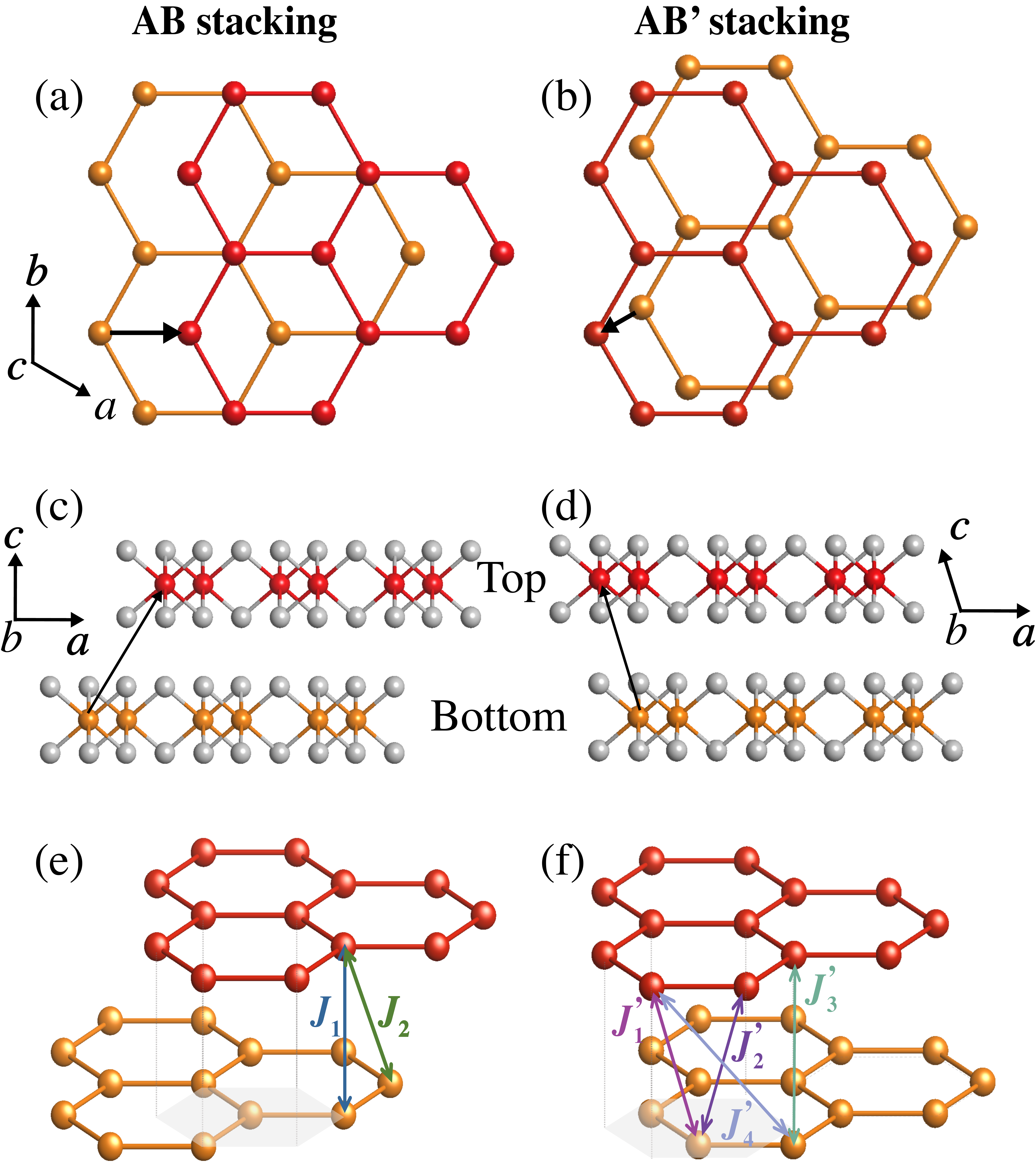}
}
\caption{\label{fig:model}
(a, b) Top views and (c, d) side views of atomic structure in bilayer VI$_3$ in AB and AB$'$ stacking. The red and orange hexagons represents honeycomb structure of V atoms in top and bottom layers, and gray balls represent I atoms. The black arrow indicates the vector which connects equivalent atoms located in two layers and shows how the top layer is sliding with respect to the bottom layer.  Inter-layer exchange coupling $J_{ij}$ in bilayer VI$_3$ for (e) AB and (f) AB$'$ stacking.
}
\end{center}
\end{figure} 

Let us now consider the case of bilayer VI$_3$, showing a similar atomic structure to its bulk counterpart. 
In particular, we studied two structures corresponding to {\it R}\=3 and {\it C}2/{\it c} phases in bulk VI$_3$ \cite{vi3_bulk3_insu}. The difference between these two structures 
is related to the different stacking  of the two VI$_3$ single layers. As shown in Fig.~\ref{fig:model} (a) and (c), the AB-stacking is characterized by the top layer showing one V atom sitting above the hexagon center of the bottom layer, similar to bilayer graphene.
Indeed, when comparing  equivalent atoms in two layers, the top layer is horizontally shifted from the bottom layers 
by $(2\bm{a} + \bm{b})/3$ (shown by a black arrow),   
where $\bm{a}$ and $\bm{b}$ denote the lattice vectors. This structure has  {\it R}\=3 symmetry. 
On the other hand, the AB$'$ stacking, with {\it C}2/{\it m} space-group symmetry, is  characterized by the shift of the top layer by $-(\bm{a}+\bm{b})/2$, as shown in Fig.~\ref{fig:model} (b) and (d).
We employed the experimental values for the in-plane lattice constants of both materials: in bulk VI$_3$, we used $a = b = 6.84$ {\AA} \cite{vi3_bulk3_insu} and for CrI$_3$, we used $a = b = 6.87$ {\AA} \cite{cri3_struc}.  
A 20 {\AA}-thick vacuum is contained in the supercell for 2D slab simulation. 

\begin{table}[!b]
\centering
\caption{ Relative total energy (meV/f.u) for   
inter-layer FM and AFM  spin configurations 
in bilayer VI$_3$ and CrI$_3$ in AB and AB$'$ stacking. The lowest energy state is highlighted. 
\label{table:ediff}}
\setlength{\tabcolsep}{13pt}
\renewcommand{\arraystretch}{1.0}
  \begin{tabular}{c c c c } 
  \hline
 &  & AB & AB$'$ \\ 
  \hline
  VI$_3$ & FM & \bf{0} & 2.44 \\
  & AFM & 0.69  & \bf{2.40} \\
 \hline
  CrI$_3$ & FM & \bf{0} & \bf{34.8} \\
  & AFM & 14.7 & 36.6 \\
 \hline
\end{tabular}
\end{table}

\begin{table*}[!hbtp]
\caption{ 
Number of equivalent bonds per unit cell $N$, bond distance between transion-metal sites $d$, and calculated exchange coupling constants $J_{ij}$ in AB and AB$'$ stacking for bilayer VI$_3$ and CrI$_3$. 
\label{table:jij}}
\centering
\setlength{\tabcolsep}{10pt}
\renewcommand{\arraystretch}{1.0}
\begin{tabular}{|c|c| c c c| c c c c c|} 
 \hline
   & &  \multicolumn{3}{|c|}{AB stacking} &   \multicolumn{5}{|c|}{AB$'$ stacking} \\
   & &  $J_{\parallel}$ &  $J_1$ & $J_2$ &  $J_{ \parallel }$ & $J'_1$ & $J'_2$ & $J'_3$ & $J'_4$ \\ 
 \hline
   VI$_3$ & $N$ & 6 & 1 & 9 & 6 & 2 & 2 & 2 & 2 \\
   &$d$ (\AA) & 3.95 & 6.66 & 7.74 & 3.95 & 7.04 & 7.05 & 8.07 & 8.95 \\
 &$J_{ij}$ (meV) & -3.20 & 0.81 & -0.24 & -4.46 & 0.10 & 0.21 & 0.04 & -0.04 \\
 \hline
    CrI$_3$ &  $N$ & 6 & 1 & 9 & 6 & 2 & 2 & 2& 2  \\
   &$d$ (\AA) & 3.95 & 6.57 & 7.68 & 3.95 & 7.00 & 7.02 & 8.03 & 8.92 \\
  &$J_{ij}$ (meV) & -7.03 & -0.82 & -0.69 & -8.11 & -0.18 & -0.23 & -0.29 & 0.25 \\ 
\hline
 \end{tabular}

\end{table*}


First, we remark that the robustness of the intra-layer FM spin ordering is demonstrated by calculating the energy difference between FM and N$\rm \acute{e}$el type antiferromagnetic AFM spin configurations in VI$_3$ monolayer, {\em i.e.} $\Delta E = E_{\rm AFM} - E_{\rm FM} =$  12.8 meV/f.u. The V magnetic moment was calculated as   2.16 $\mu_{\rm B}$. 
We then address the magnetic properties of bilayers, by calculating the total energy between inter-layer
 FM and AFM orders for the two stackings in bilayer VI$_3$ and compared the results with those obtained in CrI$_3$. 
In bilayer VI$_3$, the inter-layer FM and AFM spin configurations are very close in energy; nevertheless, the FM 
order is favored in the AB-stacking, while the  AFM order is favored in AB$'$ stacking, as reported in Table \ref{table:ediff}. 
Differently, in bilayer CrI$_3$,  the FM order is favored in both AB and AB' stacking pattern. 
Noteworthy, the inter-layer FM order in AB$'$ stacking is only slightly more stable than the AFM order, the energy differences being rather sensitive to used on-site Coulomb $U$ values, therefore not allowing a direct comparison  with previous works on bilayer CrI$_3$. 
In any case, according to the energy differences reported in Table \ref{table:ediff}, the magnetic stability in bilayers VI$_3$ and in CrI$_3$ AB$'$-stacking results to be 
weak. As such, it
may lead to an easy control of the magnetism by either external electric fields or electrostatic doping.

To understand the magnetic stability, we evaluated the magnetic exchange interactions between
V atoms by fitting total energies calculated in AB and AB$'$ stacking to the Heisenberg Hamiltonian. 
Here we assume the Heisenberg Hamiltonian,          
\begin{equation}
H = \sum_{\left < i,j  \right > } J_{ij} \bm{s}_i \cdot \bm{s}_j,
\label{eq:heisenberg}
\end{equation}
where $J_{ij}$ are the isotropic Heisenberg coupling constant between spin sites $i$ and $j$ and 
$\bm{s_}i$ is the unit vector pointing to the direction of the spin at site $i$. A parallel spin (FM) configuration is favored when \(J < 0\) and an anti-parallel (AFM) spin configuration is favored when \(J > 0\). 

In addition to the intra-layer (in-plane) first nearest-neighbor coupling ($J_{\parallel}$), we thus considered inter-layer couplings ($J_1$ and $J_2$ in AB stacking; $J'_1$, $J'_2$, $J'_3$ and $J'_4$ in AB$'$ stacking) as schematically illustrated in Fig. \ref{fig:model} (e) and (f); associated atomic pairs distances are reported in Table \ref{table:jij}. 
In particular, we performed calculations to estimate 
$J'_2$, $J'_3$ and $J'_4$ in a 2$\times$1$\times$1 supercell via the four-state energy mapping method \cite{4state, 4state1, 4state2, 4stat3, 4state4}. This method allows to consider one specific pair of spins  and remove the background interactions, therefore
allowing the calculation of the inter-layer magnetic exchange coupling constants of interest. 


In Table \ref{table:jij} we report the estimated exchange coupling constants for bilayer VI$_3$ and CrI$_3$. 

For VI$_3$, the intra-layer exchange coupling favors parallel spin state, while the inter-layer coupling eventually favors parallel spin state in AB-stacking and anti-parallel spin states in AB'-stacking, (cfr Table~\ref{table:ediff}). 
In closer detail, 
in AB stacking, $J_1$ favours anti-parallel coupling ($0.81$ meV), while $J_2$ favours parallel coupling ($-0.24$ meV). Since there is one $J_1$ bond and nine $J_2$ bonds per unit cell, overall the ferromagnetic configuration is more stable. In AB$'$ stacking, both $J_1'$ and $J_2'$ favour anti-parallel coupling ($0.1$ meV and $0.21$ meV), thus dominantly contributing to the inter-layer AFM coupling stability. For CrI$_3$, the intra-layer and inter-layer exchange coupling basically favor parallel spin states in both AB stacking and AB$'$ stacking.



In order to shed light on the microscopic mechanism behind the stacking-dependent magnetic couplings, we recall the {\it``virtual hopping"} idea based on the Hubbard model \cite{khomskii_book}. 
In particular, we discuss the results in terms of the virtual inter-layer hopping of $e_g$-$t_{2g}$ states 
supported by the inter-layer M-I-I-M super-exchange effect. 
In the weak hopping limit, 
the inter-site hopping can be treated as a perturbation to the  ground state in which magnetic ordering does not affect the energy. 
When the hopping process is allowed between occupied and unoccupied states, it in turn contributes to the ground state energy through the second-order contribution as the effective exchange energy $J_{\rm eff}=2{t}^2/U$ with hopping integral $t$ and Coulomb repulsion $U$,  the process being called ``virtual hopping". 
If we consider the direct hopping between occupied and unoccupied $3d$ states at the transition metal sites, the
parallel-spin configuration is favored if 
 the hopping is strong between majority- and majority-spin states; on the other hand, the
 anti-parallel spin configuration is favored if the hopping  between majority- and minority-spin states is strong. 
In order to discuss the virtual hopping process, 
we extracted the hopping parameters
by using a MLWF basis set, as illustrated in Fig. \ref{fig:JT} (e) and (f).  
Note that the Wannier functions are centered at V and Cr sites and spreading the tail to I sites, so that our virtual hopping process implicitly includes the $pd$ hybridization process. 
The same concept can be found in Anderson's original work on super-exchange interaction \cite{anderson1959}. 



\begin{figure}[!bhtp]
\begin{center}
{
\includegraphics[width=80mm, angle=0]{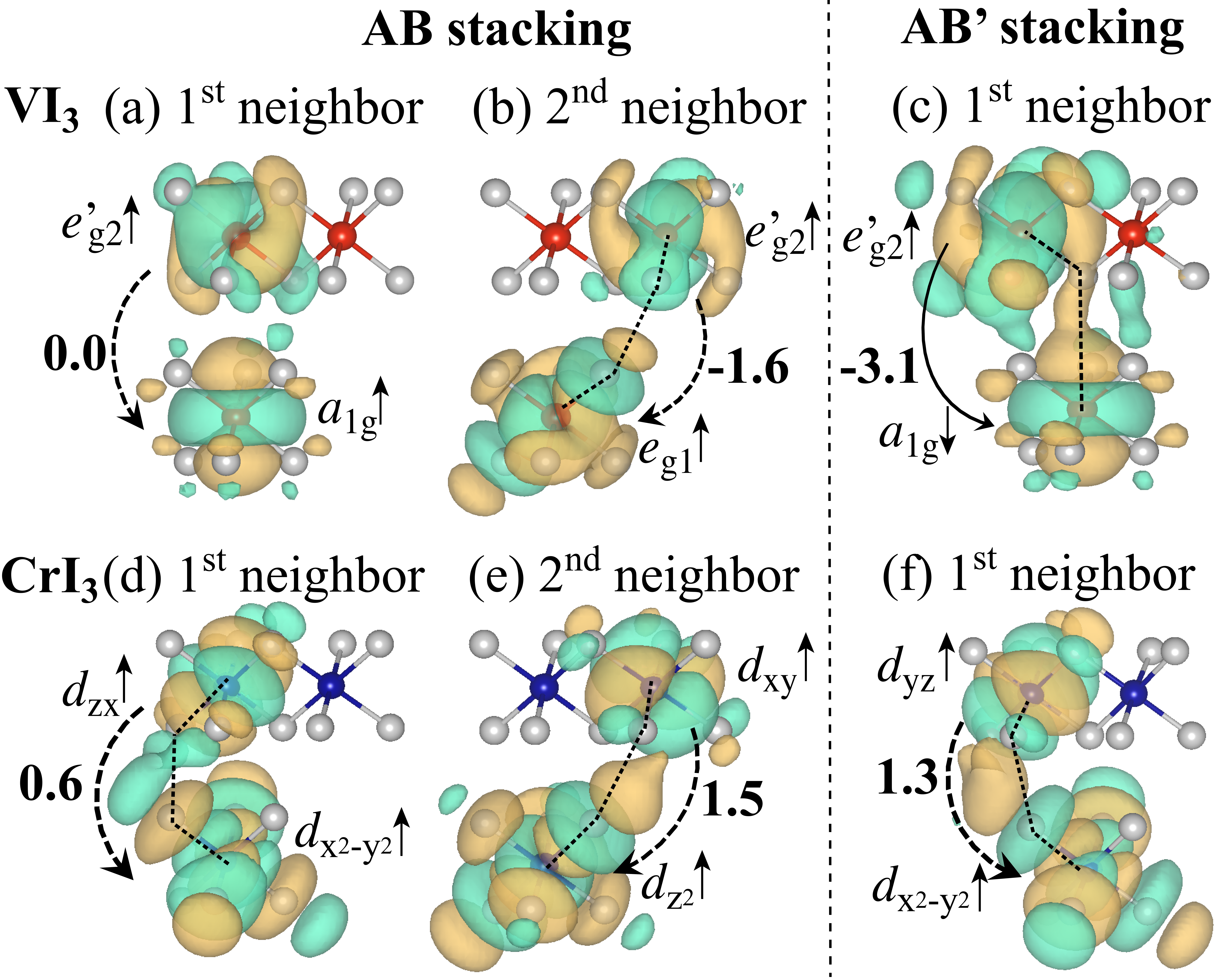}
}
\caption { \label{fig:hopping_path} 
{\color{blue}
}
MLWFs relevant to inter-layer exchange coupling in bilayer (a-c)  VI$_3$ and (d-f) CrI$_3$. 
$\uparrow$ and $\downarrow$ denote the majority and minority spin state, respectively. 
The arrows show the electron hopping from an occupied orbital state to an unoccupied orbital state; the dashed and solid lines denote the parallel-spin and anti-parallel-spin configurations, respectively. 
Values of the hopping integrals (meV) are also shown nearby the arrows. Isosurface levels were set at $\pm 0.45$ $a_0^{-3/2}$ for (a-c) and $\pm 0.3$ $a_0^{-3/2}$ for (d-f). 
}
\end{center}
\end{figure}


Figure \ref{fig:hopping_path} shows the inter-layer hopping paths with the corresponding MLWFs which are responsible for the exchange energy in bilayer VI$_3$ and CrI$_3$. 
Here we select three  types of inter-layer exchange couplings:  
first neighbor and second neighbor ($J_1$ and $J_2$) interactions in AB stacking; first neighbor ($J_1'$) interaction in AB' stacking.  The calculated hopping integrals corresponding to these exchange couplings are shown in \blue{Table. \ref{table:hopping}}. 
%


 In AB-stacking bilayer VI$_3$, 
 the trigonal CEF levels and the two-electron occupation make $J_1$ positive (anti-parallel-spin-favored). 
The hopping between $e_{g2}'^{\uparrow}$ and $a_{1g}^{\uparrow}$ states are calculated to be negligible ($t \sim$0.0~meV), thus not contributing to the magnetic interaction. 
On the other hand, the
hopping between $e_{g2}'^{\uparrow}$ and $e_{g}^{\downarrow}$ states is sizable ($t$=1.0 meV), which may be responsible for the anti-parallel-spin-favored exchange interaction. 
In AB-stacking bilayer CrI$_3$, 
in contrast, the negative (parallel-spin-favored) $J_1$ can be explained by a sizeable hopping between occupied $d_{zx}^{\uparrow}$ state and  unoccupied $d_{x^2-y^2}^{\uparrow}$ state ($t$=0.6 meV).
This is  consistent with  previous works, claiming that the $e_g$-$t_{2g}$ hopping leads to the FM coupling \cite{Jang_2019, cri3_stack1}. 
As shown in Fig.~\ref{fig:hopping_path}(d), the diagonally elongated lobes of  $d_{zx}$ and $d_{x^2-y^2}$ orbitals show a path through Cr-I-I-Cr sites with 
$d_{zx}^\uparrow$-$p$--$p$-$d_{x^2-y^2}^\uparrow$ hybridization, where the first $pd$ hybridization shows $\pi$-like and the second shows $\sigma$-like bonding. 
The second neighbor interaction $J_2$ is negative both for VI$_3$ and CrI$_3$. 
This can be explained by a large hopping integral between $e_{g2}'^\uparrow$ and $e_{g1}^\uparrow$ and that between $d_{xy}^\uparrow$ and $d_{z^{2}}^\uparrow$ states. 
In Fig.~\ref{fig:hopping_path}(b), we can recognize a $e_{g2}'^\uparrow$-$p$-$e_{g1}^\uparrow$ hybridization, where the $pd$ hybridization shows $\sigma$ bonding. For the CrI$_3$ case, a similar picture holds (cfr Fig.~\ref{fig:hopping_path}(e)), where the $e_{g2}'$ orbital is replaced by  $d_{xy}$ orbital. 

\begin{figure}[th!]
\begin{center}
{
\includegraphics[width=70mm, angle=0]{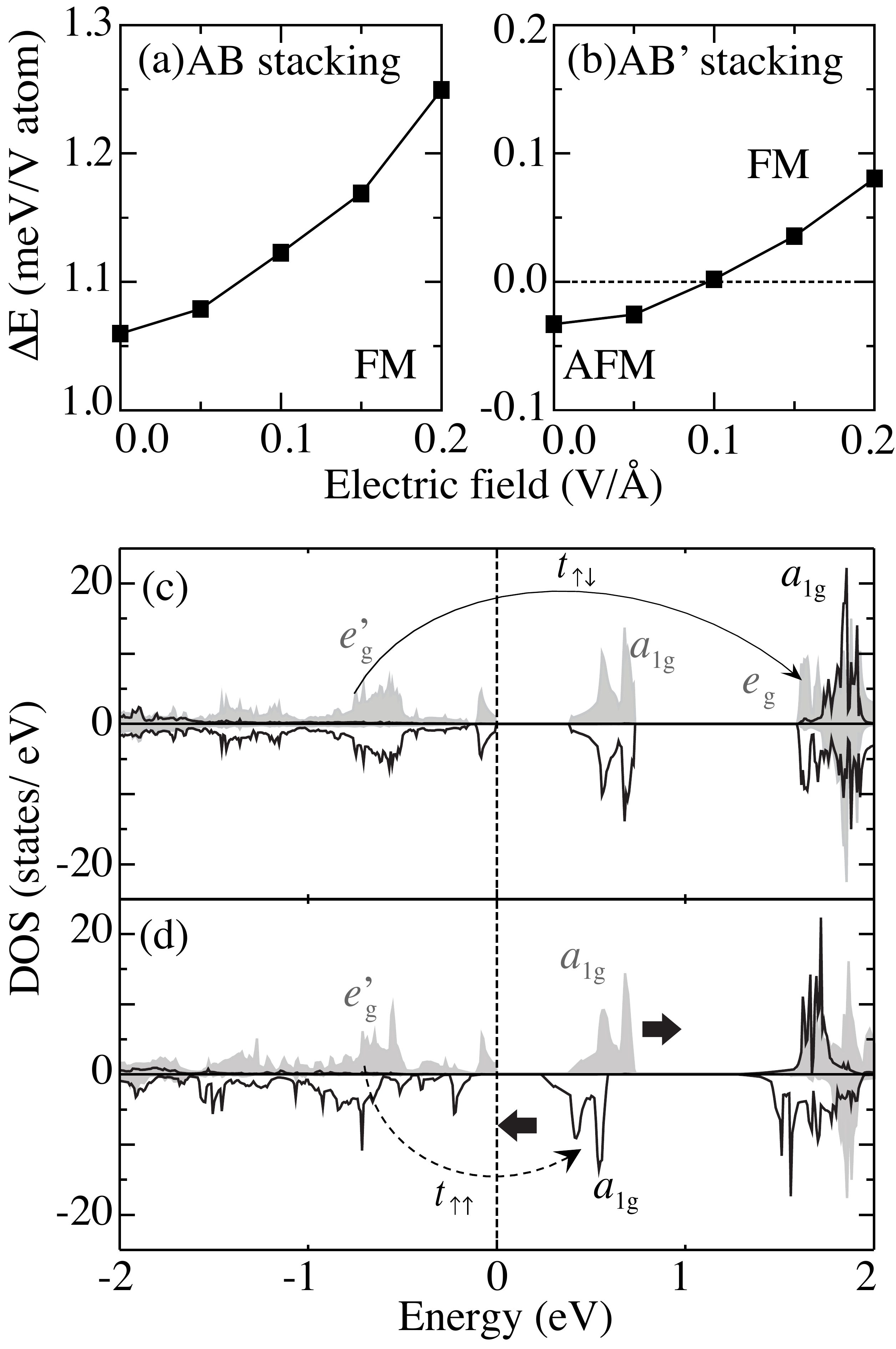}
}
\caption{\label{fig:efield}
The energy difference between the FM and AFM ordering for (a) AB and (b) AB$'$ stacking as a function of an external electric field. The positive value of $\Delta$E means FM is favored and negative means AMF is favored. $\it{d}$-orbital projected density of state for top (solid filled) and bottom (solid line) layer of bilayer VI$_3$ with AFM AB$'$ stacking in AFM ordering (c) without and (d) with external electric field. Black arrow presents for the energy shifted by applied electric field. Vertical dash line denote the Fermi energy.
}
\end{center}
\end{figure} 

The AB'-stacked bilayer VI$_3$ shows an interesting interplay between the atomic arrangement and the specific character of the $a_{1g}$ orbital of V, related to the VI$_3$ trigonal CEF:
since a vanadium atom in the bottom layer is located right under an iodine atom in the upper layer, the
V-$a_{1g}$ orbital strongly overlaps with the I-$p_z$ orbital and form the $\sigma$ bonding (Fig.~\ref{fig:hopping_path}(c)). 
This makes the $e_{g2}'$-$a_{1g}$ hopping relevant; in particular, the $e_{g2}'^\uparrow$-$p$--$a_{1g}^\downarrow$ hopping 
is very strong ($t=-3.1$ meV), making $J'_1$ positive. 
In AB' stacked bilayers, the inter-layer exchange interactions are weaker than those in AB stacked bilayers, since several possible hoppings between multiple orbital states tend to cancel each other due to the atomic arrangement.  
In evident contrast, AB'-stacked bilayer CrI$_3$ shows  results similar to the AB-stacked case: 
the $d_{yz}^\uparrow$-$p$--$p$-$d_{x^2-y^2}^\uparrow$ hybridization path, where the former (latter) $pd$ hybridization shows $\pi$ ($\sigma$) bonding, makes $J'_1$ negative (Fig.~\ref{fig:hopping_path}(f)). 
\begin{table*}[t]
\centering
\caption{ 
Hopping integrals calculated by MLWF basis set between occupied and unoccupied $d$ orbital states in parallel- ($t_{\uparrow \uparrow}$) or anti-parallel ($t_{\uparrow \downarrow}$) spin configurations. 
Three types of hopping integrals, $t_{1}$, $t_{2}$, and $t'_{1}$, corresponding with inter-layer exchange couplings $J_{1}$, $J_{2}$, and $J'_{1}$ are listed. 
$\Delta \varepsilon$  
labels the difference between two eigenenergy for the MLWFs relevant to the hopping process, corresponding to $\Delta \varepsilon_{\uparrow\uparrow}$ and $\Delta \varepsilon_{\uparrow\downarrow}$ in the main text.
The dominant hopping values relevant to the exchange couplings and those illustrated in Fig.~\ref{fig:hopping_path} are highlighted in bold. 
\label{table:hopping}}
\centering
\renewcommand{\arraystretch}{1.0}
\begin{tabular}{|c|c c c c | c c c c c |} 
  \hline
   VI$_3$  &  \multicolumn{4}{|c|}{Hopping $t_{\uparrow \uparrow}$} &   \multicolumn{5}{|c|}{Hopping $t_{\uparrow \downarrow}$}  \\
& $a_{1g}{}^0$-$a_{1g}{}^0$ &  $e_{g}{}^0$-$e_{g}{}^0$ & $e'_{g}{}^2$-$a_{1g}{}^0$ & $e'_{g}{}^2$-$e_{g}{}^0$ & $a_{1g}{}^0$-$a_{1g}{}^0$ & $e_{g}{}^0$-$e_{g}{}^0$ &$e'_{g}{}^2$-$a_{1g}{}^0$ & $e'_{g}{}^2$-$e_{g}{}^0$ & $e'_{g}{}^2$-$e'_{g}{}^0$   \\ 
  \hline
   $\Delta \varepsilon$ (eV) &  0 & 0 & 1.4 & 1.5 & 1.4 & 2.1 & 2.9 & 3.4 & 3.5 \\
 \hline
 $t_{\rm 1}$ (meV) & -3.6 & 2.1 & { \bf 0.0} & -0.7  & -4.6 & 2.7 & 0.0 & { 1.0} & { 0.6} \\
 $t_{\rm 2}$ (meV) & -0.7 & 1.1 &  0.3 & {\bf -1.6} & -0.5 & -0.9 &  -0.6 & { 2.6} & -1.0 \\
 $t'_{\rm 1}$ (meV) & 3.6 & 1.0 & -1.1 & 1.6 & 4.9 & 1.1 & { \bf -3.1} & { 3.1} & { -3.3} \\
 \hline
\end{tabular}
 \begin{tabular}{| c | c c c | c c c |} 
  \hline
   CrI$_3$ &  \multicolumn{3}{|c|}{Hopping $t_{\uparrow \uparrow}$} &   \multicolumn{3}{|c|}{Hopping $t_{\uparrow \downarrow}$}  \\
    & $e_g^0$-$e_g^0$ & $t_{2g}^3$-$e_{g}^0$ & $t_{2g}^3$-$t_{2g}^3$ & $e_g^0$-$e_g^0$ & $t_{2g}^3$-$e_{g}^0$ & $t_{2g}^3$-$t_{2g}^0$   \\ 
  \hline
    $\Delta \varepsilon$ (eV) & 0 & 1.5 & 0 & 2.7 & 4.3 & 4.6 \\
 \hline
 $t_{\rm 1}$ (meV)& 0.9  & { \bf 0.6} & 0.7 & 1.3  &  -1.1 & 1.7 \\
 $t_{\rm 2}$ (meV) &  0.6 & { \bf 1.5} & 0.8 & 0.6 & { 2.7} & -1.3 \\
 $t'_{\rm 1}$ (meV) & 2.1 & { \bf 1.3} & 1.4 & 3.0 & -3.1 & { 4.6}\\
 \hline
\end{tabular}
\end{table*}

\subsection{Electric field control of magnetic stability}

Finally, we discuss the effect of an applied electric field on the magnetic stability in bilayer VI$_3$. 
This is indeed relevant, since a magnetic phase transition upon electric-field application has been  reported in bilayer CrI$_3$ \cite{cri3_field1, cri3_field2, cri3_field3}. 
The energy difference between inter-layer AFM and FM states in AB and AB' stacked with applied electric fields is shown in Fig. \ref{fig:efield}. In both stacking cases, an applied electric field  promotes the FM ordering. Remarkably, in the AB$'$ stacking,  the ground state switches from AFM to FM ordering when the electric fields exceed a threshold value of $\sim$ 0.1 V/{\AA}. 

The microscopic mechanism of the tunable magnetic stability can be explained by invoking again the virtual hopping idea.  
The DOS projected onto V-{\it d} orbital state of top and bottom layers in  AB$'$ stacked bilayer VI$_3$ is shown in Fig. \ref{fig:efield} (c) and (d). 
Without electric field, the DOS relative to the top and bottom layer lie in the same energy range.
As discussed above, there is a competition between parallel-spin hopping and anti-parallel-spin hopping in determining the first-neighbor exchange coupling $J'_1$. 
Since the energy difference between  $e_{g}'^\uparrow$ and $a_{1g}^\downarrow$ state ($\Delta \varepsilon_{\uparrow\downarrow}=2.9$ eV) is much larger than the energy difference between $e_{g}'^\uparrow$ and 
$a_{1g}^\uparrow$ state ($\Delta \varepsilon_{\uparrow\uparrow}$=1.4 eV), one may think that a parallel-spin configuration is favored. 
However, $J'_1$ is found to be slightly AFM-favored. This is because the anti-parallel spin hopping
($t_{\uparrow\downarrow}=-3.1$ meV) is stronger than the parallel-spin hopping 
($t_{\uparrow\uparrow}=-1.1$ meV), resulting in a stronger AFM effective exchange coupling $J_{\rm eff} \propto t^2/\Delta \varepsilon$. 
Upon electric fields, the $a_{1g}$ orbital state of the top layer is shifted up, while it is shifted down in the bottom layer (cfr Fig.~\ref{fig:efield}). The band gap becomes narrower due to the shift of DOS and in turn decreases the difference of orbital energy levels, 
while the hopping integral is not significantly affected. 
Overall, this increases the tendency toward FM stability, eventually switching the favored magnetic configuration from AFM to FM. 


\section{Conclusions}
By means of first-principles calculations, we investigated the magnetic stability in bilayer VI$_3$ and compared our results with the corresponding  well-studied case of CrI$_3$. 
In particular, the magnetic exchange interactions have been analyzed by evaluating the hopping integrals between MLWFs projected onto $3d$ orbital states at V and Cr sites.
We found out that the trigonal crystal field associated to the relevant JT distortion within a single-layer of VI$_3$ (and absent in CrI$_3$), plays an important role for the inter-layer magnetic exchange interaction. 
The $t_{2g}$ orbital states are in fact split into $e_{g}'$ and $a_{1g}$ states; the latter shows the typical lobe shape pointing along the out-of-plane direction and the strong hopping between bottom-layer $a_{1g}$ and top-layer  $e_{g}'$ states determine the antiferromagnetic inter-layer coupling in AB' stacking bilayer VI$_3$. Nevertheless, since the hoppings favoring parallel-spin and anti-parallel-spin configuration are highly competing, the application of electric fields allows the switching of the inter-layer magnetic ordering from AFM to FM, paving the way to spintronic applications of  VI$_3$-based 2D magnets.



\section*{Acknowledgments}
This work was supported by 
``Center for Spintronics Research Network'', Osaka University. The numerical computation was performed on the Supercomputing Facilities at the Institute for Solid State Physics, University of Tokyo. 
KY also acknowledges Center for Computational Materials Science, Institute for Materials Research, Tohoku University for the use of MASAMUNE-IMR (Project No.20K0045). SP and DA acknowledge
financial support from the Italian Ministry for Research and Education through the Progetto Internazionale
Nanoscience Foundry and Fine Analysis (NFFA-MIUR)  
facility and through the PRIN-
2017 project ``TWEET: Towards Ferroelectricity in
two dimensions" (IT-MIUR Grant No. 2017YCTB59).

\section*{References}
\nocite{*}
\bibliographystyle{apsrev4-1}
\bibliography{mybibitem.bib}

\end{document}